\documentclass[aps,amsmath,amssymb,groupedaddress,twocolumn]{revtex4}
\usepackage{graphicx}
\usepackage{epsfig}

\newcommand{\nc}{\newcommand}
\nc{\beq}{\begin{equation}}
\nc{\eeq}{\end{equation}}
\nc{\beqa}{\begin{eqnarray}}
\nc{\eeqa}{\end{eqnarray}}

\def\gsim{\mathrel{\rlap{\lower4pt\hbox{\hskip1pt$\sim$}}
    \raise1pt\hbox{$>$}}}       

\begin{document}

\title{Life on moduli space?}

\author{Stephen~D.~H.~Hsu} \email{hsu@uoregon.edu}
\affiliation{Institute of Theoretical Science, University of Oregon,
Eugene, OR 97403 USA}

\begin{abstract}
While the number of metastable landscape vacua in string theory is vast, the number of supermoduli vacua which lead to distinct low energy physics is even larger, perhaps infinitely so. From the anthropic perspective it is therefore important to understand whether complex life is possible on moduli space -- i.e., in low energy effective theories with 1. exact supersymmetry and  2. some massless multiplets (moduli). Unless life is essentially impossible on moduli space as a consequence of these characteristics, anthropic reasoning in string theory suggests that the overwhelming majority of sentient beings would observe 1-2. We investigate whether 1 and 2 are by themselves automatically inimical to life and conclude, tentatively, that they are not. In particular, we describe moduli scenarios in which complex life seems possible.
\end{abstract}


\date{\today}

\maketitle

\bigskip

Assuming our current understanding of string theory is correct, the number of distinct vacua with unbroken supersymmetry and exact low-energy moduli (supermoduli) is infinitely larger even than the vast number of metastable (i.e., flux stabilized) string landscape vacua in which supersymmetry is broken and the cosmological constant nonzero \cite{BDG,DK}. For example, in Calabi-Yau compactifications, the continuous parameters determining the shape of the compact space are themselves moduli and result in an infinite set of physically distinct vacua. Indeed, the highly supersymmetric vacua may be on stronger theoretical footing than their non-supersymmetric counterparts \cite{BDG}.

If complex life is possible on even a tiny fraction of points on supermoduli space, it would be difficult to understand, within an anthropic framework, why we do not ourselves observe unbroken supersymmetry and massless moduli fields.

There is thus ample motivation to investigate whether complex life can exist on moduli space -- specifically, in low energy effective theories with 1. exact supersymmetry and 2. some massless multiplets (moduli). In fact property 1 will play a much larger role in our analysis than 2, because massless moduli by themselves do not seem to have any automatically disastrous consequences for life. Indeed, we have massless degrees of freedom in our universe such as the photon, and perhaps even some (fermionic) neutrino species. Further, there is no requirement that the massless multiplets be strongly coupled to the degrees of freedom from which life is made -- their interactions might be extremely weak. Strictly speaking, the string vacua which motivate this discussion are described at energies below the string scale by supergravity models. However, as complex life is likely an even lower energy phenomenon, the distinction between global and local supersymmetry (e.g., whether there is a massless gravitino, etc.) does not play an important role in our analysis.

Because of the overwhelming numerical dominance of supermoduli vacua over metastable landscape vacua, it seems reasonable to consider the subset of moduli vacua with somewhat favorable properties for the existence of life, thereby allowing us to tune model parameters such as particle masses, symmetry groups, flavor structure. That is, in the anthropic context we are comparing two quantities:
$$
{\cal N}_{\rm mod} \cdot P({\rm life \vert mod})
$$
versus
$$
{\cal N}_{\rm mls} \cdot P({\rm life \vert mls})~~,
$$
where ``mod'' refers to moduli and ``mls'' to metastable landscape vacua, ${\cal N}$ is the number of such vacua, and $P$ the conditional probability of life. If ${\cal N}_{\rm mod} \gg {\cal N}_{\rm mls}$, we can restrict our discussion to correspondingly rare (favorable) subsets of moduli vacua while still ensuring that the beings in such universes are much more typical than the ones that experience broken SUSY and no massless moduli. It is currently believed that ${\cal N}_{\rm mod}$ is uncountably infinite, whereas estimates of 
${\cal N}_{\rm mls}$ tend to be large (e.g., $10^{500}$) but finite \cite{DK}.

Note that if there exists {\it any} point in moduli space which is favorable for life, then, by continuity, there must exist a neighborhood around that point (in the many-dimensional space of moduli parameters) which is also favorable to life. This neighborhood contains an (uncountably) infinite number of distinct universes, each of which is favorable to life.

\bigskip
{\bf Desiderata for complex life}
\bigskip

Below we list some minimal requirements for complex life. In fact, we do not know whether any of these conditions are necessary or sufficient for life, although it seems they are more likely to be necessary (especially B.) than sufficient. These requirements primarily place constraints on low energy physics. As we discuss below, they do not seem to exclude moduli vacua, at least not in any obvious way.  

\bigskip
A. structure formation

\bigskip

B. deviation from thermal equilibrium (long lived sources of free energy)

\bigskip
C. stable bulk matter, complex chemistry
\bigskip

Because inflationary dynamics are typically determined by high energy physics, it seems reasonable to assume that the specific properties of any inflationary epoch (including the spectrum of density perturbations) are independent of the low energy properties of a particular vacuum, such as whether supersymmetry is broken. Therefore, the requirement of an inflationary epoch neither favors nor disfavors properties 1-2 relative to SUSY breaking at low energies. However, it is possible that inflation is much more likely in universes where SUSY is broken at high scales, e.g., near the Planck scale.

Under a volume-weighted anthropic measure (i.e., which takes into account e-folds of inflation), moduli vacua seem even more favored, since ultimately all metastable landscape vacua will eventually decay to moduli vacua \cite{eternalinflation}.

\bigskip
{\bf Supersymmetric bulk matter}
\bigskip

We begin by considering requirement C. What is the nature of supersymmetric bulk matter? Is it stable? Can it support complex chemistry? \cite{Clavelli}  

Upon first inspection, it appears that SUSY matter comes in degenerate supermultiplets. (By ``matter'' we refer specifically to {\it bulk} matter, or bound states of many SUSY particles. Presumably, complex life requires stable or long-lived metastable bound states of this type; otherwise, creatures would simply disperse into free particles.) Consider a multiparticle energy eigenstate
\begin{equation}
\label{many}
\vert \psi_1 \, \psi_2 \cdots \psi_n \rangle~~.
\end{equation}
Because the Hamiltonian is assumed to commute with the SUSY generators $Q_\eta$, all states 
\begin{equation}
Q_\eta \, \vert \psi_1 \, \psi_2 \cdots \psi_n \rangle
\end{equation}
are degenerate in energy. However, the transformed state  
\begin{equation}
\label{super}
 \vert (Q_\eta \psi_1) \, \psi_2 \cdots \psi_n \rangle ~+~ 
\vert  \psi_1 \, (Q_\eta \psi_2) \cdots \psi_n \rangle ~+~ \cdots
\end{equation}
is a superposition of components that can quickly decohere from each other once interactions with the surrounding environment are taken into account \cite{D,X}. The SUSY rotation has changed the spin and statistics of a different particle in each component, leading to distinct entanglements of each component with environmental degrees of freedom. If, for example, the state in (\ref{many}) described the hydrogen atom ground state, then (\ref{super}) would be a superposition of states with an electron in the s orbital and a spinless selectron in the s orbital. Because of the fragility of such superpositions to decoherence, observers may not, depending on the detailed dynamics which determine the pointer basis, observe states like (\ref{super}), any more than we might detect ordinary molecules in superpositions of different chiralities  \cite{decoherence}. In the case of chiral molecules an eigenstate of chirality (i.e., left- or right-handed) is always observed, even though the chiral eigenstate is a superposition of energy eigenstates. 

SUSY does {\it not} imply that the individual components in (\ref{super}) are necessarily energy eigenstates, and even if they were the first might have a different energy eigenvalue than the second. The same applies to other properties of each individual component: they could, e.g., have different electromagnetic moments, leading to different interactions with the environment and resulting in decoherence. Thus, macroscopic objects in SUSY worlds do not necessarily exhibit degeneracies. 

We now turn to the issues of stability and chemistry. For simplicity, let us consider a version of SQED and the atom-like objects which appear in such models. Electrons can rapidly convert to selectrons by emission of a photino, and selectrons in higher orbitals can reduce their energy by occupying the most tightly bound state, as the exclusion principle does not apply to bosons. Thus, one might suspect that in SUSY atoms selectrons are bound in the smallest possible s orbitals. Such atoms can only bond via van der Waals interactions, making life-supporting chemistry difficult \cite{Clavelli}. 

However, the situation is even more complex than described above. {\it SUSY matter cannot be analyzed simply in terms of individual atomic configurations.} 

Rigorous results due to Dyson and Lenard and to Lieb \cite{stability} show that non-relativistic bulk matter comprised of charged nuclei and oppositely charged bosons (i.e., bosonic electrons) interacting via two-body potentials (e.g., the Coulomb potential) is unstable -- its energy is unbounded from below.  Indeed, stability of ordinary matter built from atoms is inextricably tied to the exclusion principle obeyed by electrons, so substituting selectrons for electrons is dangerous. According to the results of Dyson et al., bulk matter formed initially from the SUSY atoms described above can lower its energy dramatically by rearranging the selectron wavefunction. What is the resulting ground state configuration?

The rigorous results assume that the interactions between charged particles can be represented by two body potentials. Instability results from shrinking the spatial size (region of support) of the many-boson wavefunction about each nucleus. However at large densities one cannot neglect $n$-body interactions and a relativistic many-body description becomes necessary. Unfortunately, rigorous results are then lacking. 

Were the results of \cite{stability} to be fully applicable to SUSY matter they would provide a very powerful anthropic argument for why we do not find ourselves on a moduli vacuum. 

However, one can argue {\it specifically} in the SUSY case that bulk matter is likely to be stable. Consider the energy per particle $E/N$ as a function of the number of particles $N$ in SUSY matter. (Assume $N$ represents some conserved quantity, analogous to baryon number; consider the ground state configuration for each value of $N$.) For bosonic matter, under the two body approximation which leads to the Hamiltonians in \cite{stability}, one would find that $E/N$ is unbounded from below. However, in a model with exact SUSY we know that the Hamiltonian is bounded from below. Therefore, we expect $E/N$ to approach some (non-zero) limiting value at large $N$. This suggests the existence of stable matter in these models, {\it albeit of extremely complex nature}: the stability of SUSY bulk matter is due to the inapplicability (breakdown) of the non-relativistic potential models studied in \cite{stability}. Whether SUSY matter has the properties to support complex life is beyond our ability to predict, and almost certainly depends on the specifics of particle masses and interactions.

Note one cannot evade these issues by postulating cosmological segregation of fermionic from bosonic particles. Any atoms formed (in the familiar way) out of fermions can easily decay to the bosonic type discussed above, by emission of gauginos. As discussed, bosonic bulk SUSY matter has lower energy than fermionic bulk matter.

\bigskip
{\bf Black holes}
\bigskip

We now turn to requirements A and B. Below we demonstrate that if black holes of appropriate mass exist, they can act as long lived sources of free energy (i.e., playing the role of stars in our universe), and as gravitational potentials which allow non-relativistic objects to decouple from cosmological expansion. 

Considering black holes allows us to avoid the usual complicated discussion of star formation. Recall again that any scenario operative on moduli space, as long as it is not overwhelmingly improbable, can dominate scenarios which operate only on metastable landscape vacua. 

Many mechanisms exist for the cosmological production of black holes, including primordial density perturbations and cosmological phase transitions \cite{primordial}. None of these mechanisms seem to be excluded by exact SUSY or massless moduli.

Black holes as sources of free energy: consider a rock of size $D$ gravitationally bound to a black hole of mass $M$. Can a black hole provide the energy required for life on the rock? How long can this last, and how many degrees of freedom can the rock support? Since our goal is to examine black holes as sources of energy, we assume the surrounding universe to be much colder than the hole, and mostly empty (other than the rock itself).

Let $\epsilon$ be the energy scale of chemistry (e.g., the typical binding energy of complex matter). In our universe $\epsilon$ is determined by atomic physics and would be of order an electron volt. We use natural units in which $\hbar = c = M_* = 1$, where $M_*$ is the Planck energy, or scale of quantum gravity.

The lifetime of a black hole of mass $M$ in $\epsilon$ units is
\beq
L \sim M^3 \epsilon^{-1}~~,
\eeq 
whereas the age of our universe in $\epsilon$ units is $L \sim 10^{32}$.

A black hole radiates energy at the rate $\dot{M} \sim M^{-2}$. Assume that the rock absorbs a fraction of this energy and achieves an equilibrium temperature of order $\epsilon$, so that the free energy is easily usable in chemical processes. Equating the black hole radiance and the energy loss rate of the rock $\sim \epsilon^4 D^2$ we obtain
\beq
M^2 \sim \frac{1}{\epsilon^4 D^2}~~.
\eeq
Let the number of chemical degrees of freedom on and in the rock be $N \sim \epsilon^3 D^3$. We adopt $N \sim 10^{50}$ as a reasonable number of degrees of freedom; roughly equivalent to the number of electrons on earth.

Then $N \sim (M \epsilon)^{-3}$, and, combining with our expression for $L$, we obtain $NL \sim \epsilon^{-4}$ or 
\beq
\epsilon \sim (NL)^{-1/4}~~.
\eeq
For, e.g., $\epsilon \sim 10^{-20}$ we can obtain $N \sim 10^{50}$ and 
$L \sim 10^{30}$. $10^{-20}$ in Planck units is 100 MeV. Working backwards, using $L \epsilon \sim M^3$, we obtain $M \sim 10^3$ ($10^{22}$ GeV), and using
$D^3 \sim N \epsilon^{-3} \sim 10^{110}$, we obtain $D \sim 10^{36}$ (of order meters). Note the rock is much more massive than the black hole, so it is the latter which orbits the former. Finally, we see that $\epsilon^4 D^3 \ll D$, so the density of the rock can be smaller than that of a black hole.

We see that an orbiting black hole can easily heat a large rock to a temperature that is amenable to life, and maintain this condition for a time (measured in units of the timescale of chemical reactions) equivalent to the age of our own universe. This constitutes an existence proof that star formation, nuclear burning, etc. are not the only mechanisms, using known physics, for generating a persistent source of free energy.

\bigskip
{\bf Supersymmetric creatures and orbiting black holes?}
\bigskip

We do not advocate that the existence of an infinite number of moduli vacua in string theory leads to a multiverse populated by an infinite variety of SUSY civilizations warmed by black holes. In fact, our ability to analyze anthropic scenarios is quite limited. In the case at hand, we have little understanding of the dynamics of complex systems formed of supersymmetric matter. Perhaps some aspect of supersymmetric matter (e.g., its detailed chemistry) makes life impossible for any choice of model parameters. To the extent that one accepts string theory and its landscape, it would seem that there is strong motivation to further investigate this topic.

\bigskip

\emph{Acknowledgments---} The author would like to thank C. Burgess, X. Calmet and D. Reeb for useful discussions. This work is supported in part by the
Department of Energy under DE-FG02-96ER40969.







\begin{thebibliography}{99}
\bibitem{eternalinflation} In some scenarios such as inflation with a first order phase transition (``old inflation''), the transition never completes and most of the volume of the universe inflates eternally. In such a scenario a volume-weighted anthropic measure would favor de Sitter space. However, such universes are cold and anisotropic -- they do not resemble ours.
\bibitem{X}
A simpler example: in a system with rotation invariance, a continuous family of superpositions of two different orientations of a molecule would all be degenerate, but interactions with the environment (e.g., scattering with soft photons) would quickly decohere the different orientation states from each other. This is a kind of symmetry breaking caused by the dynamics of the system-environment interaction. (The dynamics specify the decoherent pointer basis.) In the SUSY case the rotation is a rotation in superspace, and mixes particles with different spin and statistics. Thus, it seems quite likely that environmental interactions would distinguish the different components in the superposition. It is worth noting that supersymmetry is a fragile symmetry -- for example, if the environment is at finite temperature the distributions of bosons and fermions will already be very different.
\end{thebibliography}
\end{document}